\documentclass[11pt,a4paper]{article}
\usepackage{jcappub}

\usepackage{bm}
\usepackage{amsfonts}
\usepackage{subfigure}

\newcommand{\alt}{\mbox{\;\raisebox{.3ex}
  {$<$}$\!\!\!\!\!$\raisebox{-.9ex}{$\sim$}\;}}
\newcommand{\agt}{\mbox{\;\raisebox{.3ex}
  {$>$}$\!\!\!\!\!$\raisebox{-.9ex}{$\sim$}}\;}

\renewcommand\({\left(}
\renewcommand\){\right)}

\newcommand{\be}{\begin{equation}}
\newcommand{\ee}{\end{equation}}
\newcommand{\bea}{\begin{eqnarray}}
\newcommand{\eea}{\end{eqnarray}}

\begin{document}


\subheader{\hfill Preprint MPP-2013-113}

\title{Axion hot dark matter bounds\\ after Planck}

\author[a]{Maria Archidiacono,}
\author[a]{Steen Hannestad,}
\author[b]{Alessandro Mirizzi,}
\author[c]{Georg Raffelt}
\author[d]{and Yvonne Y.Y.~Wong}

\affiliation[a]{Department of Physics and Astronomy, University of Aarhus \\
DK-8000 Aarhus C, Denmark}

\affiliation[b]{II.~Institut f\"ur Theoretische Physik, Universit\"at
Hamburg\\ Luruper Chaussee 149, D-22761 Hamburg, Germany}

\affiliation[c]{Max-Planck-Institut f\"ur Physik
(Werner-Heisenberg-Institut)\\
 F\"ohringer Ring 6, D-80805 M\"unchen, Germany}

\affiliation[d]{School of Physics, The University of New South Wales \\
Sydney NSW 2052, Australia}

\emailAdd{archi@phys.au.dk, sth@phys.au.dk, alessandro.mirizzi@desy.de}
 \emailAdd{raffelt@mpp.mpg.de, yvonne.y.wong@unsw.edu.au}

\abstract{
We use cosmological observations in the post-Planck era to derive limits on thermally
produced cosmological axions. In the early universe such axions contribute
to the radiation density and later to the hot dark matter fraction.
We find an upper limit $m_a<0.67$~eV at 95\% C.L.\ after marginalising
over the unknown neutrino masses, using CMB temperature and polarisation data from Planck and WMAP respectively,
the halo matter
power spectrum extracted from SDSS-DR7,
and the local Hubble expansion rate $H_0$ released by the Carnegie Hubble Program based on
a recalibration of the Hubble Space Telescope Key Project sample.
Leaving out the local $H_0$ measurement relaxes the limit somewhat to 0.86~eV, while Planck+WMAP alone constrain the axion mass to 1.01~eV, the first time an upper limit on $m_a$ has been obtained from CMB data alone. Our axion limit is therefore not very sensitive to the tension between the Planck-inferred $H_0$ and the locally measured value.
This is in contrast with the upper limit on the neutrino mass sum, which we find here to range from $\Sigma \, m_\nu < 0.27$~eV at 95\% C.L.\ combining all of the aforementioned observations, to $0.84$~eV from CMB data alone.}

\maketitle

\section{Introduction}

Observations of the large-scale structure in the universe, most notably
of the cosmic microwave background (CMB) anisotropies and the galaxy distribution, are a
powerful probe of both the expansion history and the energy content
of the universe. The precision of these observations is such that in some cases
cosmological constraints on particle physics models that could plausibly account for some of the
energy content are more stringent than limits obtained from direct laboratory measurements. One notable example is the
cosmological bound on the neutrino mass which is currently close to
an order of magnitude more restrictive than existing laboratory
bounds \cite{Lesgourgues:2006nd,Hannestad:2010kz,Wong:2011ip,Lesgourgues:2012uu}. Given this
ability of cosmology to probe neutrino physics, it is of significant
interest to investigate how well other hot dark matter (HDM) models with
similar behaviours can be tested.

One example of particular interest is the case of hot dark matter
axions. In reference~\cite{Hannestad:2003ye} some of the present authors found that
cosmological data do in fact  provide an upper bound on the mass of
hadronic axions comparable to that for  neutrinos. The analysis was
subsequently updated on several occasions to include progressively more recent data
releases~\cite{Hannestad:2005df,Hannestad:2007dd,Hannestad:2008js,Hannestad:2010yi},
again with similar conclusions.   Other authors also found comparable
results~\cite{Melchiorri:2007cd}. Here, we provide another update incorporating
 the most recent measurement of the CMB temperature anisotropies by the Planck mission~\cite{Ade:2013xsa,Ade:2013lta}, as well as
the matter power spectrum extracted from
the seventh data release of the Sloan Digital Sky Survey (SDSS-DR7)~\cite{Reid:2009xm},
and the present-day Hubble parameter obtained from the most recent recalibration of
the Hubble Space Telescope (HST) Key Project sample as part of the Carnegie Hubble Program~\cite{Freedman:2012ny}.

The exercise of extracting information about low-mass particles from precision
cosmological data has received a new twist in the  past few
years because combinations of different data sets are not necessarily
consistent with the minimal (``vanilla'') $\Lambda$CDM cosmology.
In particular, many pre-Planck observations prefer at the $2 \sigma$ level a ``dark radiation'' component equivalent to an additional
$\sim 0.5$ massless neutrino species (see, e.g., \cite{Hou:2012xq,Archidiacono:2013lva,Calabrese:2013jyk,Abazajian:2012ys}
and references therein).
The Planck data alone and in combination with baryon acoustic oscillations
do not support the presence of a dark radiation component.  However, the preferred value of the Hubble parameter $H_0$ is across the board significantly lower than that from direct measurements in our local neighbourhood~\cite{Freedman:2012ny}; if these measurements were to be brought in line with one another, then the presence of a dark radiation equivalent to $\sim 0.6 \pm 0.25$ could achieve the required result through its degeneracy with $H_0$~\cite{Ade:2013lta}.  However, hovering around the $2 \sigma$ level the case for dark radiation, pre-Planck or post-Planck, is certainly not strong.   Indeed, it could be as much a signature for new physics as it is a signature for poorly controlled systematics in some cosmological data sets.  Nonetheless, since we are dealing here with low-mass particles which do contribute to the primordial radiation density,
some additional care is required when dealing with local~$H_0$ measurements.

The rest of the paper is organised as follows. In section~\ref{sec:axions} we briefly discuss the hot dark matter axion scenario.  We describe the cosmological parameter space to be analysed in section~\ref{sec:model}, while section~\ref{sec:data} summarises the data sets.  Section~\ref{sec:results} contains our results, notably,  bounds on the axion mass $m_a$ and the neutrino mass sum $ \Sigma \, m_\nu$ from various data combinations, as well as a brief discussion of the impact of hot dark matter on the CMB observables and the corresponding degeneracies with $H_0$.  We conclude in section~\ref{sec:conclusions}, followed by an appendix~\ref{sec:appendix}  in which we discuss the degeneracies of $m_a$ and $\Sigma \, m_\nu$ in more detail.

\section{Hot dark matter axions}                    \label{sec:axions}

The Peccei--Quinn solution of the CP problem of strong interactions
predicts the existence of axions, low-mass pseudoscalars that are
very similar to neutral pions, except that their mass and interaction
strengths are suppressed relative to the pion case
by a factor of order $f_\pi/f_a$, where
$f_\pi\approx 93$~MeV is the pion decay constant, and $f_a$ is a large
energy scale known as the axion decay constant or Peccei--Quinn
scale~\cite{Peccei:2006as,Kim:2008hd,Kawasaki:2013ae}. Specifically,
 the axion mass~is
\begin{equation}\label{eq:axmass}
 m_a=\frac{z^{1/2}}{1+z}\,\frac{f_\pi}{f_a} m_\pi
 =\frac{6.0~{\rm eV}}{f_a/10^6~{\rm GeV}}\,,
\end{equation}
where $z=m_u/m_d$ is the mass ratio of the up and down quarks. We
shall follow the previous axion literature and assume the canonical
value $z=0.56$. We note that variation in the range
0.38--0.58 is possible~\cite{Beringer:1900zz}, although this uncertainty has no
strong impact on our discussion. Henceforth we shall mostly use $m_a$
as our primary phenomenological axion parameter, instead of the more
fundamental parameter~$f_a$.

Axions with masses in the $10~\mu{\rm eV}$ region are well motivated
as a cold dark matter (CDM) candidate, albeit with considerable uncertainty
as regards the precise mass value~\cite{Kawasaki:2013ae,Sikivie:2006ni,Wantz:2009it,Hiramatsu:2012gg}.
Such CDM axions would have been produced non-thermally by the
re-alignment mechanism and, depending on the cosmological scenario,
by the decay of axion strings and domain walls. In addition, a sizeable hot
axion population can be produced by thermal processes if the Peccei--Quinn scale
is sufficiently low~\cite{Chang:1993gm,Turner:1986tb,Masso:2002np,Graf:2010tv}.  For $f_a\alt10^8$~GeV, or, equivalently, $m_a\agt60~{\rm meV}$,
axions attain thermal equilibrium at the QCD phase transition or later, and contribute to the cosmic radiation density and subsequently
to the cosmic hot dark matter  along with massive neutrinos.

In principle, the parameter region $f_a\alt 10^9$~GeV ($m_a\agt10~{\rm meV}$) has already been excluded
by the SN~1987A neutrino burst
duration~\cite{Mayle:1987as,Raffelt:1987yt,Turner:1987by,Raffelt:2006cw}
based on the axion--nucleon interaction. Furthermore, axions in the same parameter region
would also contribute strongly to neutron-star
cooling~\cite{Umeda:1997da,Keller:2012yr} and engender a cosmic diffuse
supernova axion background (DSAB)~\cite{Raffelt:2011ft}. However,
the sparse data sample, our poor understanding of the nuclear medium
in the supernova interior, and simple prudence suggest that one
should not base far-reaching conclusions about the existence of
axions in this parameter range on a single argument or experiment
alone.  For those axion models in which the axion couples nonvanishingly
to the electron (DFSZ-type axion models~\cite{Zhitnitsky:1980tq,Dine:1981rt}),
numerous other stellar energy loss limits based on the axion--electron coupling
also exclude axions with masses exceeding some 10~meV~\cite{Raffelt:1985nj,Raffelt:1994ry,Isern:1992,Corsico:2012ki,%
Corsico:2012sh,Isern:2008nt,Isern:2008fs,Barth:2013sma}.  We therefore
do not consider these (well-constrained) models, but rather focus on hadronic models (KSVZ-type models
\cite{Kim:1979if,Shifman:1979if}) in which the axion does not directly
couple to ordinary quarks and leptons, and in so doing bypasses constraints from stellar evolution arguments.
Hadronic axion models are also simple to analyse in that all axion properties depend on $f_a$ alone.

\begin{figure}
\centering
\begin{tabular}{c}
\includegraphics[scale=0.4]{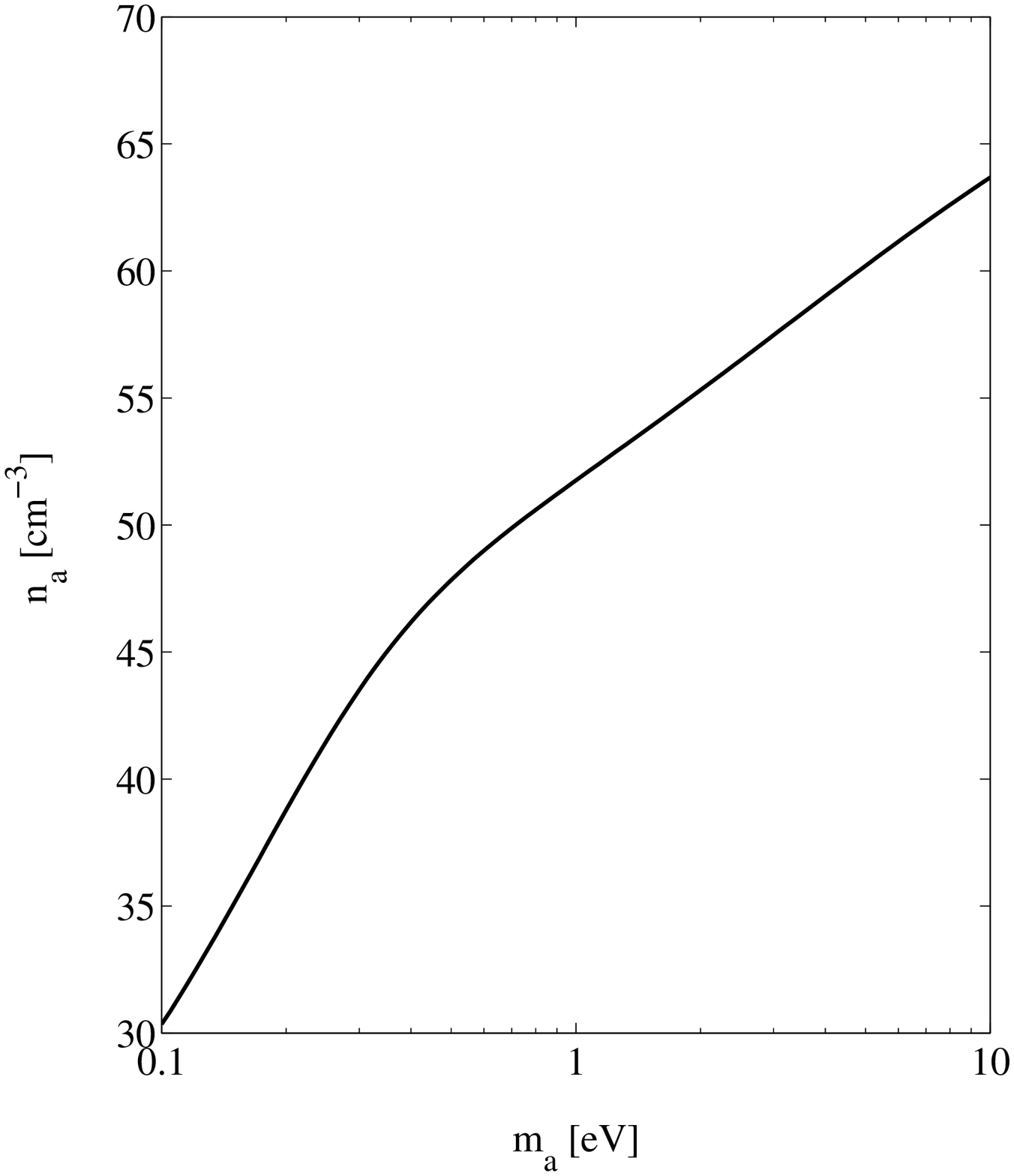}\\
\includegraphics[scale=0.4]{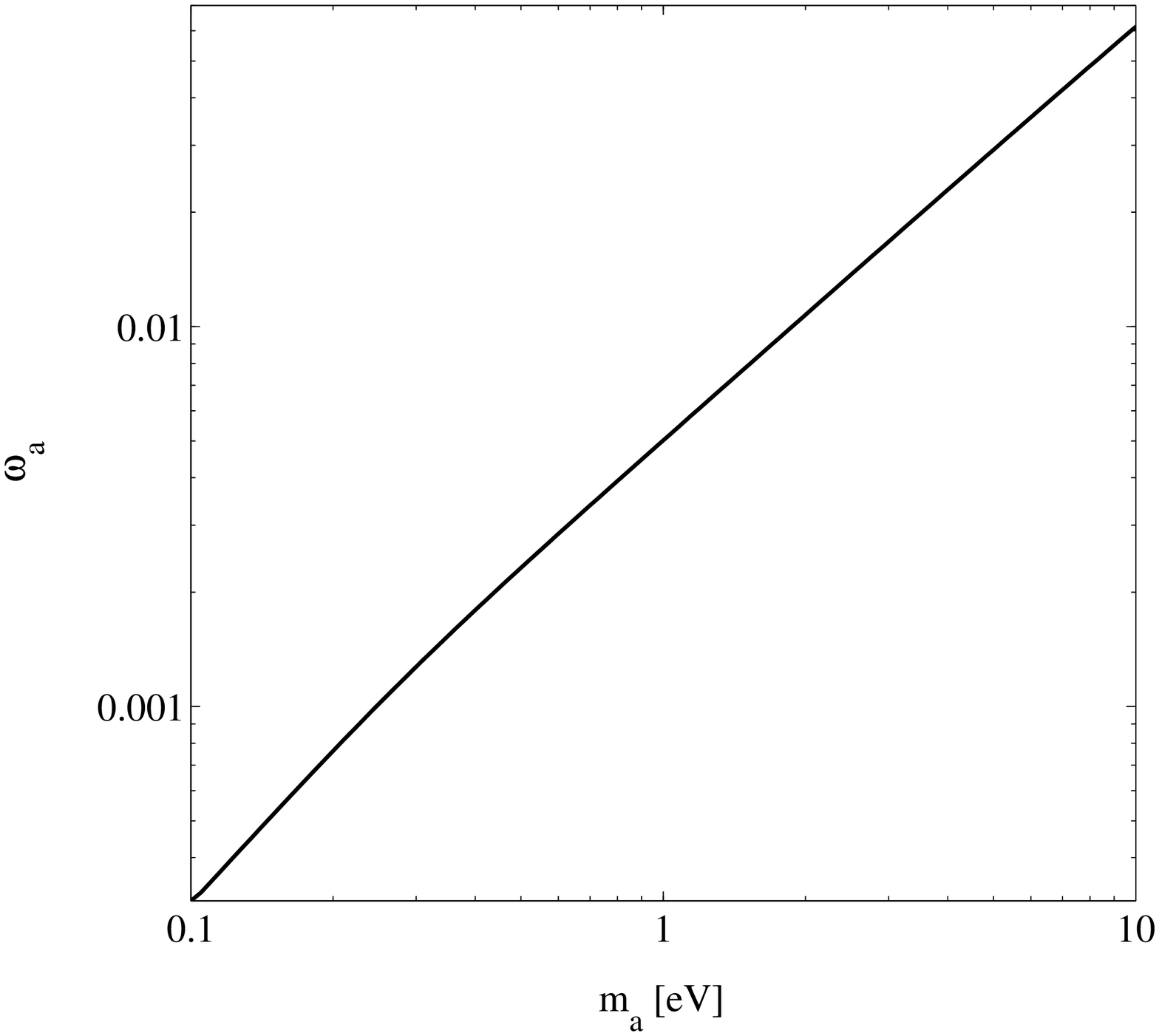}
\end{tabular}
\caption{{\it Top}: Axion number density $n_a$ as a function of the axion mass $m_a$.
{\it Bottom}: Present-day axion energy density $\omega_a$ as a function of $m_a$.}
\label{fig:nama}
\end{figure}

If axions do not couple to charged leptons, the main thermalisation
process in the post-QCD epoch is $a+\pi\leftrightarrow\pi+\pi$
\cite{Chang:1993gm}. The axion--pion interaction is given by the
Lagrangian~\cite{Chang:1993gm}
\begin{equation}\label{eq:axionpionlagrangian}
{\cal L}_{a\pi}=\frac{C_{a\pi}}{f_\pi f_a}\,
\left(\pi^0\pi^+\partial_\mu\pi^- +\pi^0\pi^-\partial_\mu\pi^+
-2\pi^+\pi^-\partial_\mu\pi^0\right)
\partial_\mu a\,,
\end{equation}
where, in hadronic axion models, the coupling constant is
$C_{a\pi}=(1-z)/[3\,(1+z)]$. Based on this interaction, the axion
decoupling temperature $T_{a,{\rm dec}}$ was calculated earlier in~\cite{Hannestad:2005df},
from which follows the present-day axion number density,
\begin{equation}
n_a=\frac{g_{*S}(T_{\rm today})}{g_{*S}(T_{a,{\rm dec}})} \times \frac{n_\gamma}{2},
\end{equation}
where $g_{*S}(T)$ denotes the effective entropy degrees of freedom, and $n_\gamma$ is the present-day photon number density.
The top panel of figure~\ref{fig:nama} shows $n_a$ as a function of the axion mass~$m_a$.
Before thermal axions become nonrelativistic by cosmological redshift, they
contribute to the total radiation density an amount
\begin{equation}
\rho_a = \frac{2}{3}\(\frac{3}{2}\frac{n_a}{n_\nu}\)^\frac{4}{3}\, \frac{7}{8}\(\frac{4}{11}\)^{\frac{4}{3}} \rho_\gamma
=
\Delta N_{\rm eff} \ \frac{7}{8}\(\frac{4}{11}\)^{\frac{4}{3}} \rho_\gamma
\label{eq:neff}
\end{equation}
equivalent to $\Delta N_{\rm eff}$ additional light neutrino species,
where $n_\nu=113~{\rm cm}^{-3}$ is the present-day neutrino number
density in one flavour, and $\rho_\gamma$ is the photon energy density.
We stress that
equation~(\ref{eq:neff}) is strictly valid only when the axion temperature far exceeds $m_a$.
For order 0.1--1~eV-mass axions this corresponds to a time well before photon decoupling;
the phenomenology of such axions in the context of the CMB anisotropies and the matter power spectrum is closely linked to the axion mass scale, and
{\it cannot} be reparameterised in terms of
$\Delta N_{\rm eff}$ {\it massless} neutrino species.
Indeed, as we shall show in section~\ref{sec:results} and appendix~\ref{sec:appendix}, the degeneracy of $m_a$ especially in the direction of the Hubble parameter is vastly different from that of $\Delta N_{\rm eff}$  massless neutrinos.

\section{Cosmological model}\label{sec:model}

We perform the analysis in the context of a
$\Lambda$CDM model extended to include hot dark matter neutrinos and axions, and
vary a total of eight parameters:
\begin{equation}\label{eq:model}
{\bm \theta} = \{\omega_{\rm cdm},\omega_{\rm b},\theta_{\rm s},\tau,
\ln(10^{10}A_{s}),n_{s},
\Sigma\,m_\nu,m_a\}.
\end{equation}
Here, $\omega_{\rm cdm} \equiv \Omega_{\rm cdm} h^2$ and $\omega_{\rm b}
\equiv \Omega_{\rm b} h^2$ are the present-day physical CDM and baryon densities
respectively, $\theta_{\rm s}$ the angular the sound horizon, $\tau$  the optical
depth to reionisation, and $\ln(10^{10}A_{s})$ and $n_s$ denote respectively the amplitude and spectral index
of the initial scalar fluctuations.  We use the  neutrino mass sum $\Sigma\,m_\nu$ and the axion mass $m_a$ as fit parameters, noting that these can be easily converted to their corresponding present-day energy density via
\begin{eqnarray}
\omega_\nu &=& \frac{\Sigma \, m_\nu}{94 \ {\rm eV}}, \label{eq:omeganu} \\
\omega_a & = & \frac{2}{3}\(\frac{m_a}{94 \ {\rm eV}}\)\(\frac{T_a}{T_\nu}\)^3, \label{eq:omegaa}
\end{eqnarray}
where $T_\nu$ and $T_a$ are the neutrino and the axion present-day temperature respectively,
and $\omega_a$ is shown in the bottom panel of figure~\ref{fig:nama} as a function of $m_a$.
Table~\ref{tab:priors} shows the priors adopted in the analysis.
We stress that $\Sigma \, m_\nu$ is always let free to vary in our analysis, because neutrino masses are an experimentally established fact, not some novel untested physics that can be discarded at will.

\begin{table}[t]
\caption{Priors for the cosmological fit parameters considered in this work. All priors are uniform (top hat) in the given intervals.}
\label{tab:priors}
\begin{center}
\begin{tabular}{lc}
\hline
 Parameter & Prior\\
\hline
$\omega_{\rm b}$ & $0.005 \to 0.1$\\
$\omega_{\rm cdm}$ & $0.001 \to 0.99$\\
$\theta_{\rm s}$ & $0.5 \to 10$\\
$\tau$ & $0.01 \to 0.8$\\
$n_{s}$ & $0.9 \to 1.1$\\
$\ln{(10^{10} A_{s})}$ & $2.7 \to 4$\\
$\Sigma\,m_\nu$ [eV] &  $0 \to 7$\\
$m_a$ [eV] &  $0 \to 7$\\
\hline
\end{tabular}
\end{center}
\end{table}

For convenience we assume the neutrino mass sum $\Sigma \, m_\nu$ to be shared equally among three standard model neutrinos.  Note however that implementing
any other mass spectrum will return essentially the same conclusions,
since current observations are primarily sensitive to $\Sigma \, m_\nu$, not to the exact mass splitting.  In fact, discerning the neutrino mass spectrum---and thereby distinguishing between the normal and the inverted neutrino mass hierarchies---is unlikely to be possible even with the extremely precise measurements
from the future {\sc Euclid} mission~\cite{Hamann:2012fe}.

\section{Data and analysis}\label{sec:data}

We consider three types of measurements: temperature and polarisation power spectra of the CMB anisotropies, the large-scale matter power spectrum, and direct measurements of the local Hubble expansion rate.  These are discussed in more detail below.  To these data sets we apply a Bayesian statistical inference analysis using the publicly available Markov Chain Monte Carlo
parameter estimation package {\sc CosmoMC}~\cite{Lewis:2002ah} coupled to the CAMB~\cite{Lewis:1999bs} Boltzmann solver modified to accommodate two hot dark matter components.  With the exception of the local Hubble parameter measurement, the likelihood routines and the associated window functions are supplied by the experimental collaborations.

\subsection{CMB anisotropies}

Our primary data set is the recent measurement of the CMB temperature (TT) power spectrum by the Planck mission~\cite{Planck:2013kta}, which we implement into our
likelihood analysis following the procedure reported in~\cite{Ade:2013lta}.
This data is supplemented by measurements of the CMB polarisation from the WMAP nine-year data release~\cite{Bennett:2012fp}, in the form of
an autocorrelation (EE) power spectrum at $2<\ell<32$ and a cross-correlation (TE) with the Planck temperature measurements in the same multipole range.  We denote this supplement ``WP''.

\subsection{The matter power spectrum}

We use the matter (halo) power spectrum (HPS) determined from the
luminous red galaxy sample of the seventh data release of the SDSS-DR7~\cite{Reid:2009xm}. The full data set
consists of 45 data points, covering wavenumbers from $k_{\rm min} =
0.02\ h {\rm Mpc}^{-1}$ to $k_{\rm max} = 0.2\ h {\rm Mpc}^{-1}$. We
fit this data set following the procedure of
reference~\cite{Reid:2009xm}, using an adequately smeared power
spectrum to model nonlinear mode-coupling constructed according to the
method described in reference~\cite{Hamann:2010pw}.

\subsection{Local Hubble parameter measurements}

We further impose a constraint on the present-day Hubble parameter based on measurements of the universal expansion rate in our local neighbourhood.  The most
recent value released as part of the Carnegie Hubble Program~\cite{Freedman:2012ny} is
\begin{equation}
H_0 = [74.3 \pm 1.5 \, {\rm (stat)}\,\pm 2.1 \, {\rm (syst)}]~{\rm km}~{\rm s}^{-1}~{\rm Mpc}^{-1},
\end{equation}
which uses observations from the Spitzer Space Telescope to calibrate the Cepheid distance scale in the HST Key Project sample.
Following the analysis of reference~\cite{Ade:2013lta}, we add the statistical and systematic errors in quadrature,
assuming both uncertainties to be Gaussian-distributed.

\section{Results and discussions}\label{sec:results}

\begin{figure}[t]
\centering
\includegraphics[scale=0.45]{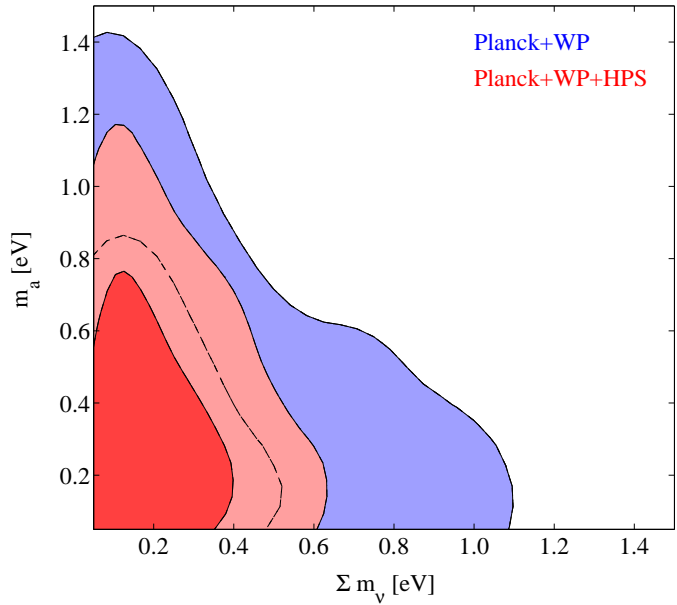}
\caption{Two-dimensional marginal 68\% and 95\% probability density contours in the $(\Sigma\,m_\nu,m_a)$-plane derived from Planck temperature and WMAP polarisation (Planck+WP)
measurements (blue shading) and in combination with the halo power spectrum (red shading).}
\label{fig:mamnu}
\end{figure}

The main results of our inference analysis are presented in table~\ref{tab:results} and figure~\ref{fig:mamnu}, which show, respectively, the
one-dimensional marginal limits on $\Sigma \, m_\nu$ and $m_a$ derived from various data combinations, and the corresponding two-dimensional marginal probability density contours in
the $(\Sigma \, m_\nu,m_a)$-subspace.

\begin{table}
\center
{
\caption{One-dimensional marginal 95\% bounds on the neutrino mass sum and the axion mass from
various data combinations.\label{tab:results}} \medskip
\centering
\begin{tabular}{lcc}
\hline
Data set  & $\Sigma\,m_\nu$ [eV] & $m_a$ [eV]\\
\hline
Planck+WP  & $ <0.84$ & $<1.01$    \\
Planck+WP+HPS   & 0.007--0.48  & $<0.86$    \\
Planck+WP+$H_0$ & $<0.25$ & $<0.78$ \\
Planck+WP+HPS+$H_0$ & $<0.27$  & $<0.67$ \\
\hline
\end{tabular}}
\end{table}

\subsection{Axion mass limit from CMB alone}\label{sec:mafromcmb}

Most strikingly, while it was previously not possible to obtain an upper limit on the axion mass from the
CMB anisotropies alone, we now find that the combination of Planck+WP is able to constrain $m_a$
quite stringently to $m_a<1.01$~eV~(95\% C.L.).  This is in contrast with the CMB-only upper limit on the neutrino mass sum $\Sigma \, m_\nu$, which has been hovering around 1~eV since the WMAP five-year data release: here, we find  $\Sigma\,m_\nu<0.78$~eV (95\% C.L.) after marginalising over the unknown axion mass,  somewhat stronger than the pre-Planck value of 1.19~eV \cite{Hannestad:2010yi}.

The fundamental and irreducible effect of 0.1--1~eV-mass hot dark matter neutrinos and axions on the CMB temperature anisotropies are similar in that both particle species
 transit from a ultrarelativistic state to a nonrelativistic one in close proximity to the photon decoupling era, as well as to the epoch of matter--radiation equality.  This transition impacts strongly on the evolution of the radiation-to-matter energy density ratio, which in turn affects the evolution of the spacetime metric perturbations and ultimately the photon temperature fluctuations.  To quantify the precise effect of this transition on the CMB temperature power spectrum naturally necessitates solving the relevant Boltzmann equations.  However, a general rule of thumb is that, assuming a fixed present-day total matter density,
those perturbation modes that enter the horizon  before the hot dark matter particle becomes fully nonrelativistic will suffer more strongly from decay of the metric perturbations in a mixed CDM+HDM cosmology than in a pure CDM scenario.  This is translated firstly into an additional ``uplift'' of the primary CMB temperature fluctuations at last scattering on the same length scales.  Secondly, the same decaying metric perturbations also alter the so-called early integrated Sachs--Wolfe (ISW) effect, which contributes to the CMB temperature anisotropies primarily around the first acoustic peak.

\begin{figure}[t]
\centering
\includegraphics[scale=0.4]{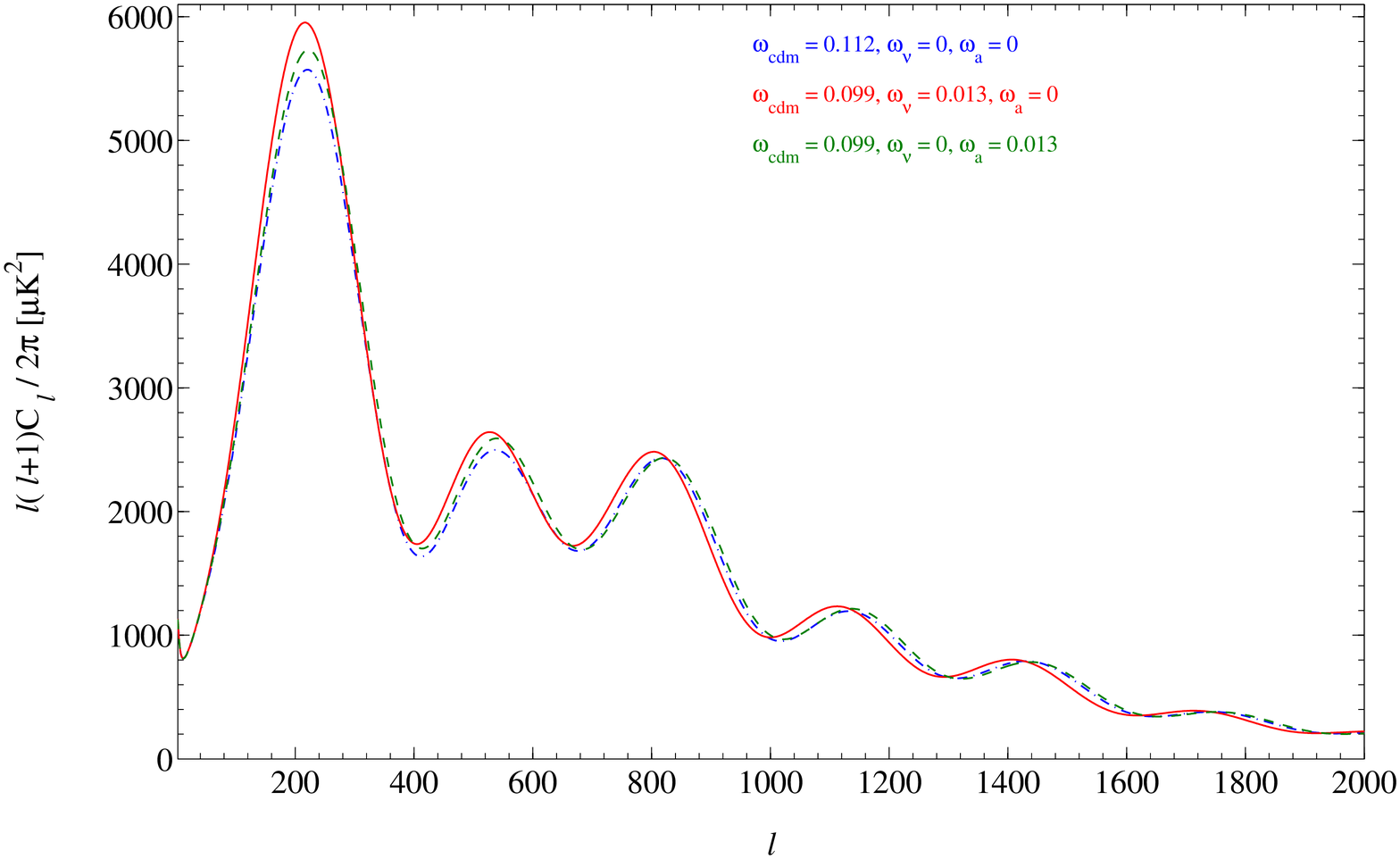}
\includegraphics[scale=0.4]{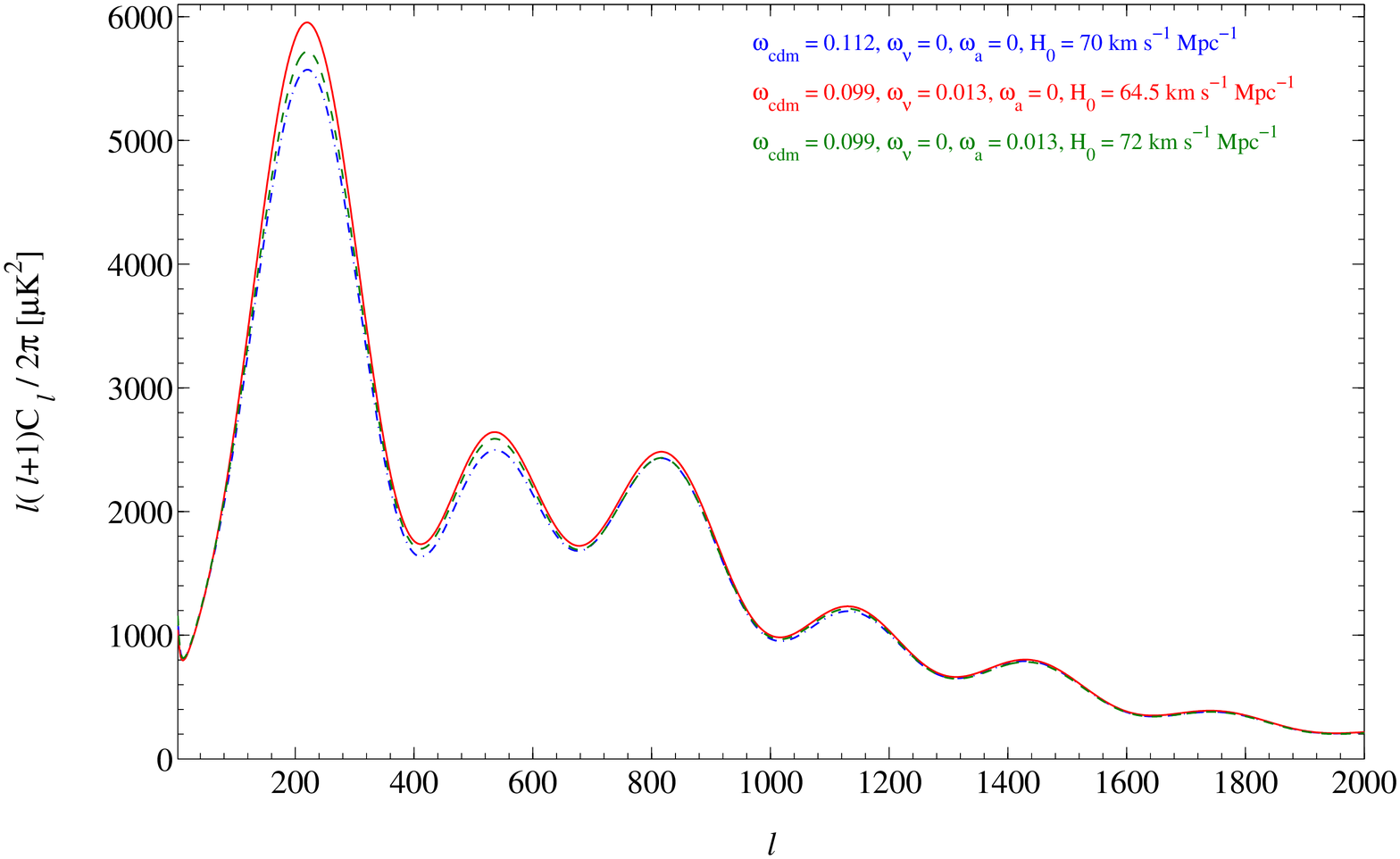}
\caption{CMB temperature angular power spectra for three different cosmological
models, all with the same total dark matter density $\omega_{\rm dm} =
\omega_{\rm cdm} + \omega_{\rm \nu} + \omega_{\rm a} = 0.112$.  The blue/dot-dash line corresponds to the reference $\Lambda$CDM model ($\omega_{\rm cdm}=0.112$), the green/dashed line an axion HDM scenario with $m_a=2.4$~eV ($\omega_{\rm cdm}=0.099$,  $\omega_a=0.013$), and the red/solid line the case of three degenerate neutrinos with masses
summing to $\Sigma \, m_\nu=1.2$~eV  ($\omega_{\rm cdm}=0.099$, $\omega_\nu=0.013$).
In the top panel we hold all other cosmological parameters fixed, while in the bottom panel we further adjust $H_0$
in order to match the peak positions---the corresponding $H_0$ values are, in units of ${\rm km} \ {\rm s}^{-1} \ {\rm Mpc}^{-1}$,
70, 72, and 64.5 for the $\Lambda$CDM, the axion and the neutrino model respectively.
}
\label{fig:cls}
\end{figure}

Following from this reasoning, what distinguishes between a hot dark matter neutrino population and an axion one of the same present-day (nonrelativistic) energy density is essentially two-fold.
Firstly, the axion population is always colder than the neutrino population by virtue of having decoupled from the cosmic plasma at an earlier time.  Being a scalar degree of freedom the axion must also have a larger particle mass than the neutrino in order to make up the same present-day energy density [see equations~(\ref{eq:omeganu})and~(\ref{eq:omegaa})].  Together these push the axion's relativistic-to-nonrelativistic transition to a higher redshift, so that its irreducible features on the CMB anisotropies occur on smaller scales (or higher multipoles).
Secondly, while the axion is a new source of energy density {\it in addition} to the energy content of the $\Lambda$CDM model, in endowing neutrinos with masses we are effectively ``stealing'' from the model's radiation content.  This effect combined with the axion's relative coldness and heaviness means that a hot axion population will always cause less deviation to the radiation-to-matter energy density ratio, and therefore leave a weaker signature on the CMB than a neutrino population of the same present-day energy density.  Figure~\ref{fig:cls} shows the CMB temperature angular power spectra for a neutrino and an axion HDM scenario, juxtaposed against their $\Lambda$CDM counterpart.

Thus what has enabled Planck to constrain the axion mass for the first time from CMB data alone can be traced to its significantly better measurement of the higher multipole moments from the third acoustic peak onwards (recall that WMAP is only cosmic-variance limited up to $\ell  \sim 500$, roughly the second acoustic peak), which rules out much of the axion parameter space in the $>1$~eV region manifesting on these scales.  In contrast, the neutrino mass could already be constrained to order 1~eV by CMB data alone
since the WMAP three-year data release \cite{Spergel:2006hy}
via the early ISW effect on the third-to-first acoustic peak height ratio; new, high multipole measurements from Planck offer no significant improvement because there are no new unique signatures from sub-eV-mass neutrinos on these scales.  See also discussions in appendix~\ref{sec:appendix}.

\subsection[Correlation with $H_0$]{Correlation with \boldmath{$H_0$}}\label{sec:correlation}

Combining CMB measurements with the halo power spectrum from SDSS DR-7 tightens the axion mass bound to
$m_a < 0.86$~eV~(95\%~C.L.), comparable to the pre-Planck value of 0.82~eV~\cite{Hannestad:2010yi}.
The neutrino mass bound likewise improves to $\Sigma \, m_\nu < 0.48$~eV~(95\%~C.L.).  Note that formally the combination of Planck+WP+HPS yields also
a  95\% lower limit of $\Sigma \, m_\nu> 0.007$~eV.  However, because neutrino oscillation experiments already limit the neutrino mass sum to $> 0.06$~eV, this cosmological lower limit is of little phenomenological consequence.

\begin{figure}[t]
\centering
\includegraphics[scale=0.45]{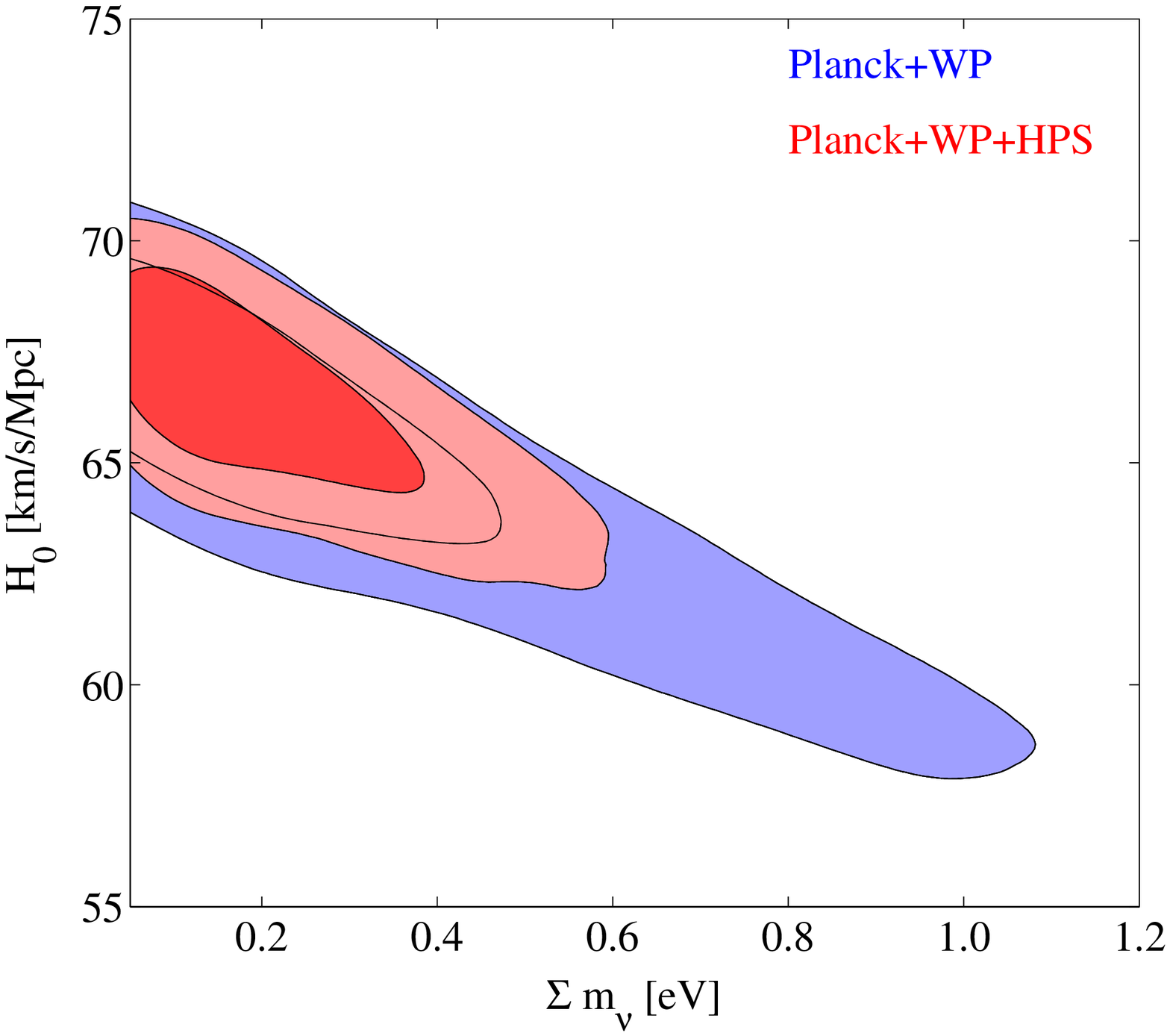}
\caption{Same as figure~\ref{fig:mamnu}, but in the $(\Sigma\,m_\nu,H_0)$-plane.}
\label{fig:mnuH0}
\end{figure}

\begin{figure}[t]
\centering
\includegraphics[scale=0.45]{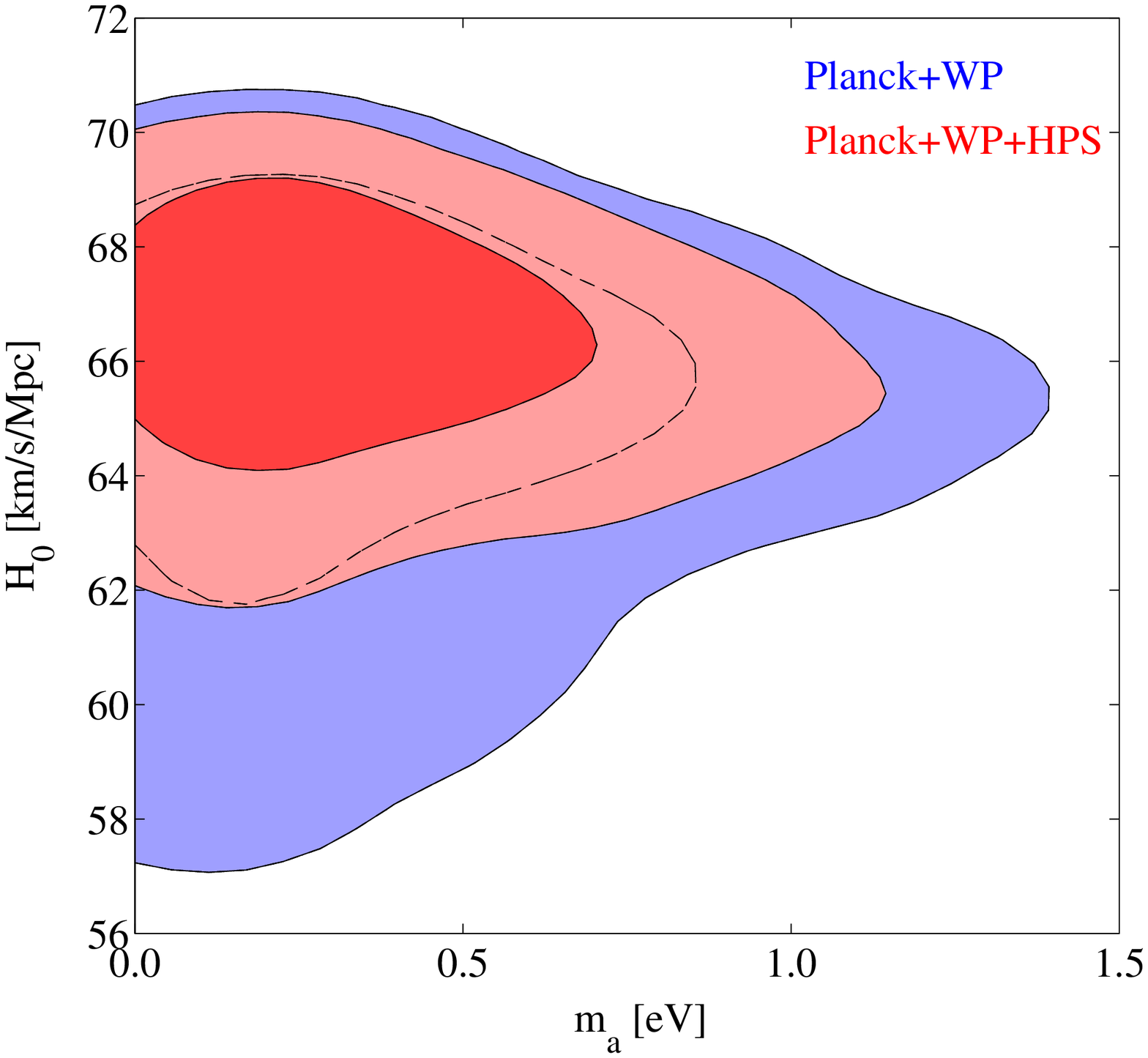}
\caption{Same as figure~\ref{fig:mamnu}, but in the $(m_a,H_0)$-plane.}
\label{fig:maH0}
\end{figure}

\begin{figure}[t]
\centering
\includegraphics[scale=0.45]{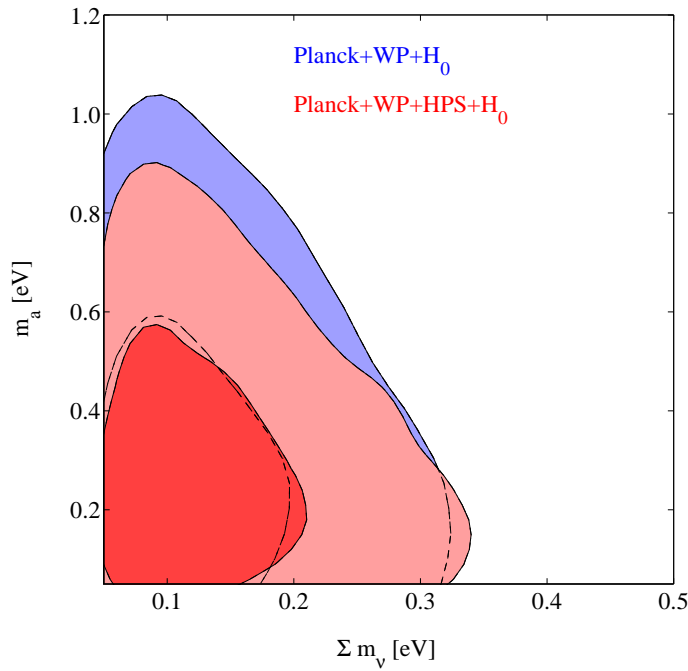}
\caption{Same as figure~\ref{fig:mamnu}, but including the $H_0$ value determined from local measurements.}
\label{fig:mamnuH0}
\end{figure}

An interesting effect arises when we consider also local measurements of the Hubble parameter $H_0$;
while the neutrino mass bound continues to tighten to $\Sigma \, m_\nu < 0.25$~eV~(95\% C.L.), inclusion of the local $H_0$ value leads only to a marginal 10\% improvement
on the axion mass bound.  This effect can be understood from an inspection of
figures~\ref{fig:mnuH0} and~\ref{fig:maH0}, which show, respectively, the two-dimensional marginal density contours in
the $(\Sigma \, m_\nu,H_0)$- and $(m_a,H_0)$-subspaces.  Clearly, a strong anti-correlation exists between $\Sigma \, m_\nu$ and $H_0$, so that adding to the analysis the local $H_0$ value---which is higher than the Planck-inferred value---tends to tighten, perhaps artificially, the neutrino mass bound.  In contrast,
no such correlation exists between $m_a$ and $H_0$, making the inference of $m_a$ independent of the value of~$H_0$.  Thus we conclude that the axion hot dark matter bound is robust against the
unresolved discrepancy between the Planck-inferred  $H_0$ value and that measured in our local neighbourhood.
Figure~\ref{fig:mamnuH0} shows the two-dimensional marginal probability density contours in the $(\Sigma \, m_\nu, m_a)$-subspace with the inclusion of local $H_0$ data.

A deeper question now is, what causes $\Sigma \, m_\nu$ and $m_a$ to exhibit different correlations with~$H_0$ in the first place.  We defer a more detailed discussion of the parameter degeneracies in the CMB temperature anisotropies to appendix~\ref{sec:appendix}, but note here that at the most basic level, any degeneracy (or lack thereof) of $\sum m_\nu$ and $m_a$ with $H_0$ can be traced to the parameter dependence of the angular sound horizon $\theta_{\rm s}  \equiv r_{\rm s} (z_\star)/D_{\rm A} (z_\star)$,
where
\begin{eqnarray}
r_{\rm s} (z_\star) &\equiv& \int^\infty_{z_\star} {\rm d} z \ \frac{c_{\rm s} (z)}{H(z)} , \label{eq:rs}\\
 D_{\rm A}(z_\star) & \equiv&\int^{z_\star}_0 \frac{{\rm d}z}{H(z)} \label{eq:da}
\end{eqnarray}
are the sound horizon at decoupling and the angular diameter distance to the last scattering surface respectively, $c_{\rm s}(z)$ is the sound speed in the tightly-coupled photon--baryon fluid, and $z_\star \simeq 1100$ is the photon decoupling redshift.  The angular sound horizon determines the location of the acoustic peaks in the CMB angular power spectrum, and is arguably the most well-measured CMB observable.

Clearly, the presence of the Hubble expansion rate $H(z)$ in the expressions~(\ref{eq:rs}) and~(\ref{eq:da}) indicates {\it a priori} a three-way degeneracy between the neutrino (or axion) energy density (and hence mass), the CDM energy density $\omega_{\rm dm}$,%
\footnote{The baryon density in principle also affects $H(z)$.  However, $\omega_{\rm b}$ can be pinned down by the CMB odd-to-even peak ratios.  We therefore consider it to be a well-measured parameter.}
and the present-day Hubble parameter~$H_0$,%
\footnote{In a flat spatial geometry, the dependence of $H(z)$ on the vacuum energy $\Omega_\Lambda$ can be reparameterised in terms of $H_0$.}
meaning that for any choice of $\Sigma \, m_\nu$ or $m_a$, we can tweak $\omega_{\rm dm}$ and/or $H_0$ to maintain the same $\theta_{\rm s}$ value.
However, the amount of tweaking required for the axion scenario is generally less than that needed by massive neutrinos.  This is because, as discussed in section~\ref{sec:mafromcmb}, an axion population generally causes less deviation to the evolution of the radiation and matter energy densities than a neutrino population of the same present-day energy density.  Indeed, assuming a fixed present-day total matter density, figure~\ref{fig:cls} shows that while the presence of a $\Sigma \, m_\nu=1.2$~eV hot neutrino population causes a noticeable shift in the peak locations that can be restored by lowering the $H_0$ value to $64.5~{\rm km} \ {\rm s}^{-1} \ {\rm Mpc}^{-1}$ (compared to
the reference $\Lambda$CDM value of $70~{\rm km} \ {\rm s}^{-1} \ {\rm Mpc}^{-1}$),
the shift due to a $m_a=2.4$~eV axion population is of a much lesser extent and requires only a small {\it positive} tuning of $H_0$ to $72~{\rm km} \ {\rm s}^{-1} \ {\rm Mpc}^{-1}$.
As we shall demonstrate in appendix~\ref{sec:appendix}, the same negative correlation with $H_0$ remains true for the neutrino case even after accounting for uncertainties in the total matter density, while for the axion scenario the small tuning of $H_0$ may be negative or positive depending on the choice of $\omega_{\rm dm}$ so that overall no correlation exists between $m_a$ and $H_0$.


\section{Summary and conclusions}\label{sec:conclusions}

We have re-examined cosmological bounds on thermal axions based on
the recent CMB temperature anisotropy measurement provided by the Planck mission, as well as other types of cosmological
observations.
For the first time CMB data alone provide a
restrictive limit on the axion mass of $m_a<1.01$~eV (95\% C.L.), which
improves to 0.86~eV with the inclusion of the SDSS matter power spectrum, and to
0.78~eV when combined with local measurements of~$H_0$.
The changes brought about by combining CMB data with other observations are obviously fairly minor, suggesting that
the systematic uncertainties in our limits are small.  In particular,
the unsettled discrepancy between the Planck-inferred $H_0$ value and that measured directly in our local neighbourhood
does not strongly affect the axion limit. This is to be contrasted with the corresponding limit on the neutrino mass sum, which sees a
dramatic reduction from $\Sigma \, m_\nu < 0.84$~eV (95\% C.L.) from CMB alone to $0.25$~eV when combined with local $H_0$ measurements.

Note that when inferring the axion mass bound, we always marginalise over the unknown
neutrino masses which contribute an unavoidable but unknown HDM fraction to the
universe; in this sense, HDM contributions by neutrinos and axions are not
alternatives, since the former must always be present.  However,  unless neutrinoless double-beta decay eventually shows a
signal, laboratory experiments are unlikely to provide neutrino mass information on the cosmologically relevant
scale in the foreseeable future.  Thus marginalisation over $\Sigma \, m_\nu$ is arguably the only consistent approach to cosmological parameter estimation, and
this situation is unlikely to change any time soon.

In comparison with the pre-Planck state of affairs, although we now have an axion mass bound
from post-Planck CMB data alone, the overall limit has not moved significantly away from order~1~eV.
This situation is of interest for solar axion searches
by the Tokyo axion helioscope~\cite{Moriyama:1998kd,Inoue:2002qy,Inoue:2008zp} and by the CAST
experiment at
CERN~\cite{Zioutas:2004hi,Andriamonje:2007ew,Arik:2008mq,Arik:2011rx}.
The axion--photon conversion efficiency in a dipole magnet directed
towards the sun strongly degrades if the photon--axion mass difference
is too large. This mass-mismatch can be overcome by filling the
conversion pipe with a buffer gas at variable pressure that provides
the photons with an adjustable refractive mass. The final CAST
search range using $^3$He as buffer gas has reached a search mass of
1.17~eV, which is hard to push up any further~\cite{Arik:2013}. In this
sense, cosmological and helioscope axion mass limits are still largely
complementary.

In the future, further improvements in HDM bounds on the cosmological front
will likely come from the next generation of large-scale galaxy and cluster surveys, notably the ESA
{\sc Euclid} mission~\cite{Laureijs:2011mu}.  {\sc Euclid} has the potential to probe neutrino masses
with a sensitivity of $\sigma(\Sigma \, m_\nu) \simeq 0.01$~eV, sufficient to measure at $5\sigma$ the minimum mass sum of $\Sigma \, m_\nu \simeq 0.05$~eV
established by neutrino flavour oscillation experiments~\cite{Hamann:2012fe}.  A similar sensitivity to HDM axions should likely be possible.

\section*{Acknowledgements}

We thank Jan Hamann for help with the initial setup of the Planck
likelihood code, and the anonymous referee for constructive suggestions that improve the clarity of the manuscript.
We acknowledge use of computing resources from the
Danish Center for Scientific Computing (DCSC), and
partial support by the Deutsche Forschungsgemeinschaft through grant
No.~EXC~153 and by the European Union through the Initial Training
Network ``Invisibles,'' grant No.\ PITN-GA-2011-289442.

\appendix

\section{Degeneracies of \boldmath{$\Sigma \, m_\nu$} and \boldmath{$m_a$} in the CMB temperature anisotropies}\label{sec:appendix}

Figures~\ref{fig:mnuH0} and~\ref{fig:maH0} show that while a strong anti-correlation exists between the neutrino mass sum $\Sigma \, m_\nu$ and the Hubble parameter $H_0$, no such correlation is present for the axion mass $m_a$ and $H_0$.  To understand these degeneracies, we note first of all that the most well-measured observable in the CMB temperature anisotropies is the angular sound horizon~$\theta_{\rm s}$, defined in equations~(\ref{eq:rs}) and~(\ref{eq:da}), which fixes the acoustic peak positions.  Assuming a fixed baryon density (as we shall do for the rest of this section), $\theta_{\rm s}$ depends on the dark matter density~$\omega_{\rm dm}$, the neutrino mass sum~$\Sigma \, m_\nu$ (or the axion mass $m_a$ as the case may be), and the Hubble parameter $H_0$.  Thus, already at this level it is clear that $H_0$ and $\Sigma \, m_\nu$ (or $m_a$) must be correlated somehow.  Indeed, as already shown in figure~\ref{fig:cls}, holding  $\omega_{\rm cdm}+\omega_\nu$ (or $\omega_{\rm cdm}+\omega_a$) fixed, the effect of $\Sigma \, m_\nu$ on the positions of the acoustic peak  can be countered by lowering the value of $H_0$, while a finite $m_a$ can be offset by raising $H_0$ by a small amount.

The odd peak height ratios are likewise a well-measured set of quantities, and are usually described in the literature as a sensitive probe of the epoch of matter--radiation equality~$a_{\rm eq}$.
A more accurate description, however, is that the ratios reflect the time evolution of the radiation-to-matter energy density ratio $f(a)$, which in the vanilla $\Lambda$CDM model (and in a few simple extensions of this model) is a one-parameter function subject to our choice of $a_{\rm eq}$ relative to the time of photon decoupling.  A simple example where this is still the case is the extension of vanilla $\Lambda$CDM with a nonstandard number of massless neutrinos,  $N_{\rm eff} \neq 3.046$.  Here, a larger $N_{\rm eff}$ can be offset by a larger matter density so as to restore $a_{\rm eq}$ and consequently {\it all} odd peak height ratios.  The degeneracy between $N_{\rm eff}$ and $\omega_{\rm dm}$ is therefore exact as far as the function $f(a)$ and its effects on the CMB are concerned,
 and when combined with the measurement of the angular sound horizon $\theta_{\rm s}$ discussed above, gives rise to the well-known positive correlation
 between $N_{\rm eff}$ and $H_0$.

The case of a nonzero  neutrino mass or axion mass is less straightforward, because no amount of $\omega_{\rm dm}$ adjustments could completely counter the effect of $\Sigma \, m_\nu$ or $m_a$ on the radiation-to-matter energy density ratio.  To begin with, no unique definition of matter--radiation equality exists in this case, because for the 0.1--1~eV-mass range under consideration there will always be some HDM particles that are neither ultra-relativistic nor completely nonrelativistic in the vicinity of photon decoupling.  Furthermore, because HDM particles are continuously losing momentum through redshift, the additional redshift dependence they engender in $f(a)$ cannot be mimicked over the whole timespan by simply tweaking $\omega_{\rm dm}$, contrary to the case of $N_{\rm eff}$.
Therefore, any correlation between $\omega_{\rm dm}$ and $\Sigma \, m_\nu$ or between $\omega_{\rm dm}$ and $m_a$ is necessarily approximate and subject to measurement errors (or cosmic variance).

\begin{figure}[t]
\centering
\includegraphics[scale=0.4]{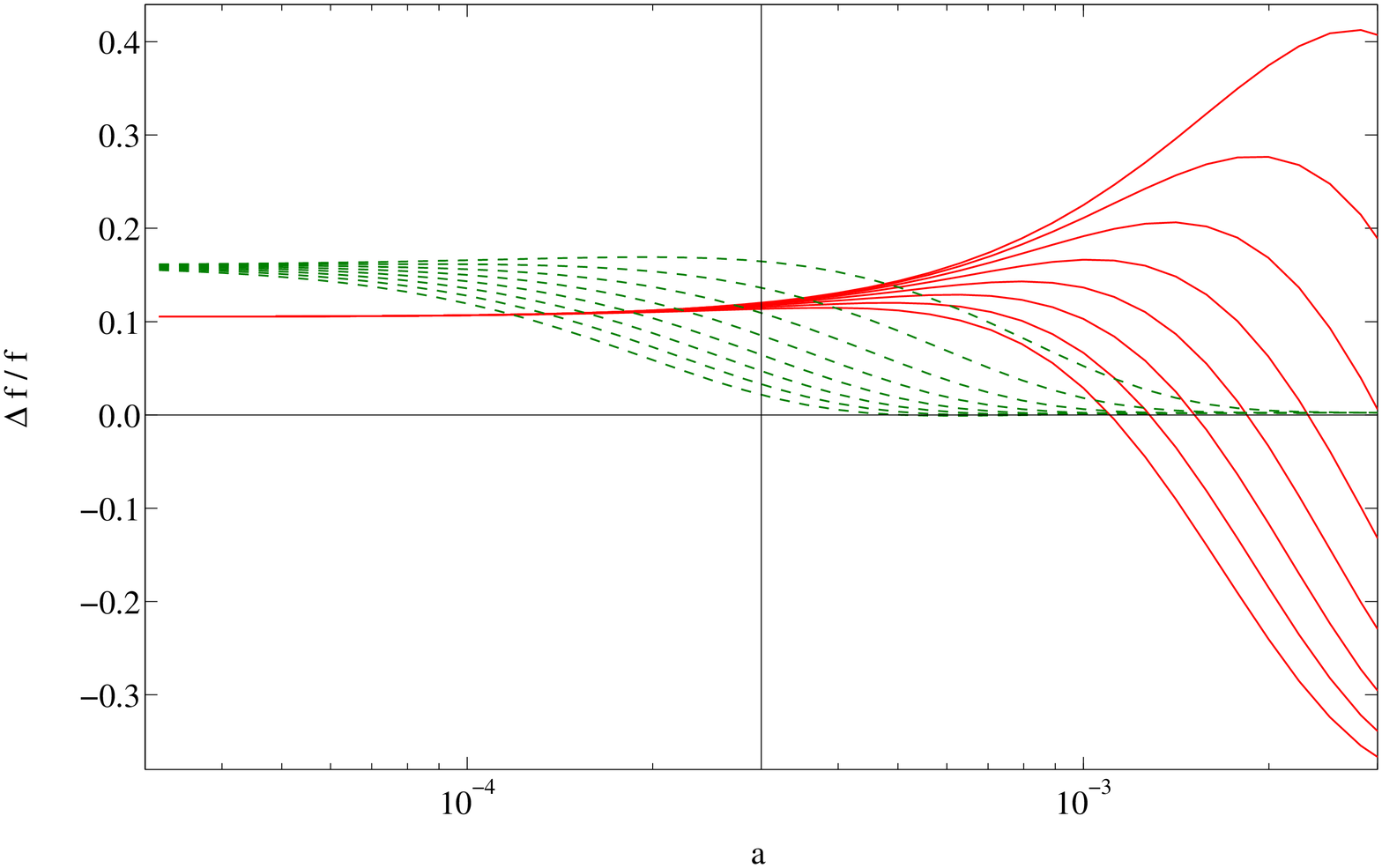}
\includegraphics[scale=0.4]{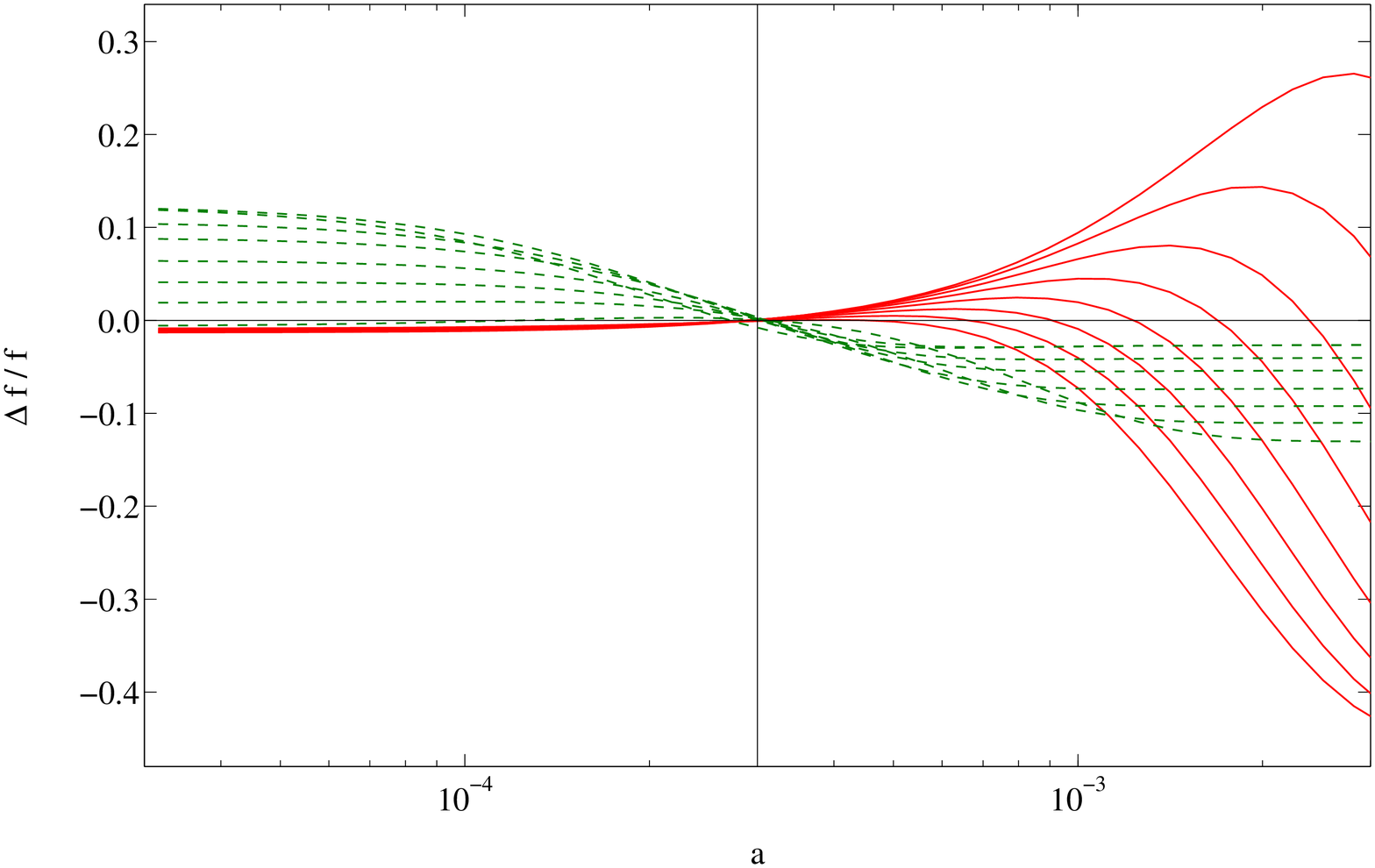}
\caption{Fractional change in the radiation-to-matter energy density ratio $f(a)$ relative to the reference $\Lambda$CDM model, as a function of the scale factor $a$.  The red/solid lines denote to the case of three degenerate neutrinos with masses
$\Sigma \, m_\nu=1.2$~eV, while the green/dashed lines represent a $m_a=2.4$~eV axion population.  A set of eight lines is shown in each case, corresponding to $c_{\rm nr}=0.3, 0.4, \ldots, 1$, where $1/c_{\rm nr}$ is defined as the ratio of the rest mass to the maximum momentum of the nonrelativistic HDM particle concerned.
In the top panel the total dark matter density is always held fixed at $\omega_{\rm dm}=0.099$. In the bottom panel we increase $\omega_{\rm dm}$ by an amount 0.014--0.0146 in the neutrino case and by 0.003--0.0205 in the axion case in order that  $f (a_{\rm eq})$, where $a_{\rm eq} \simeq 0.0003$ is the epoch of matter--radiation equality in the $\Lambda$CDM model, stays the same in all cases.}
\label{fig:rz}
\end{figure}

\begin{figure}[t]
\centering
\includegraphics[scale=0.4]{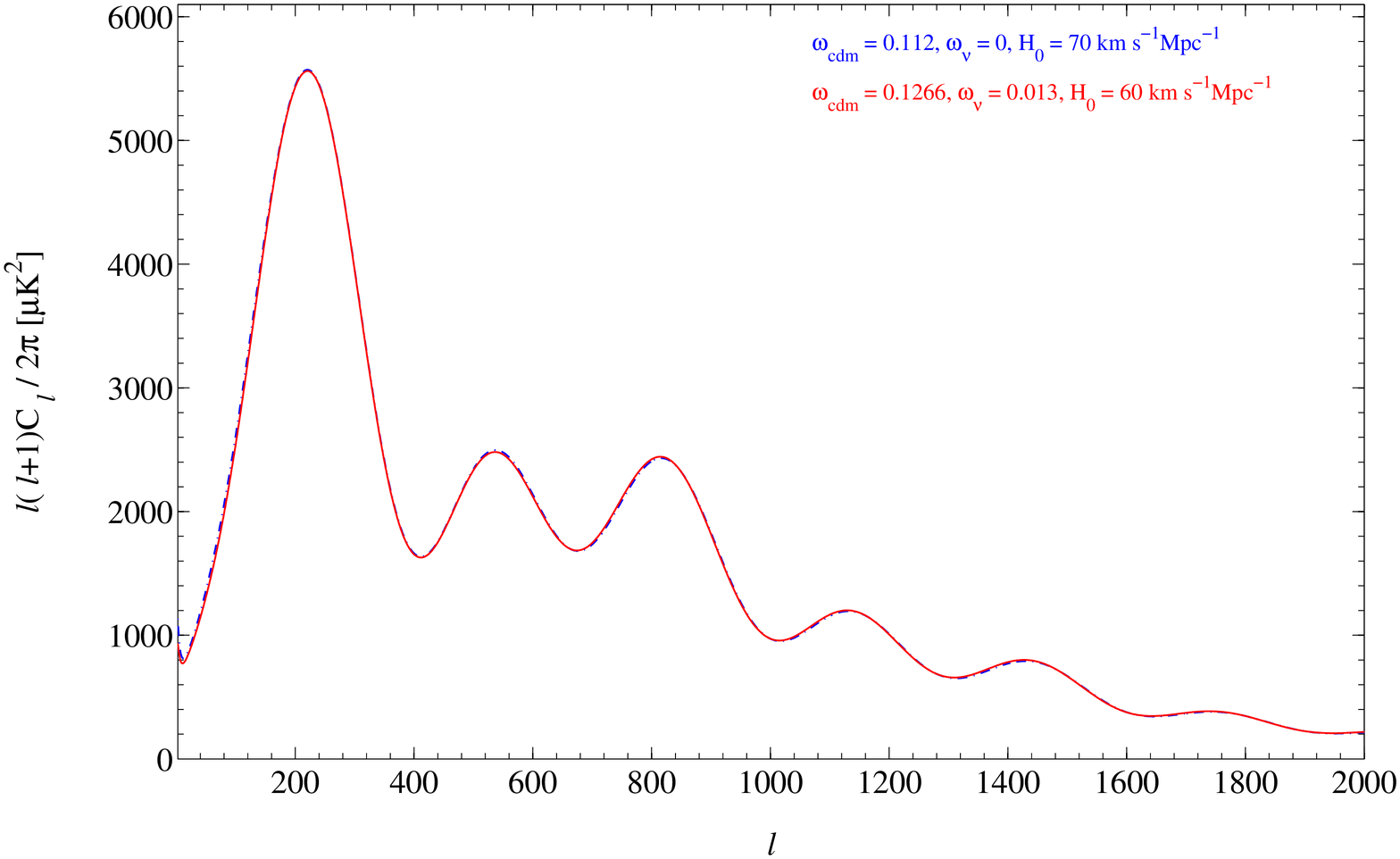}
\includegraphics[scale=0.4]{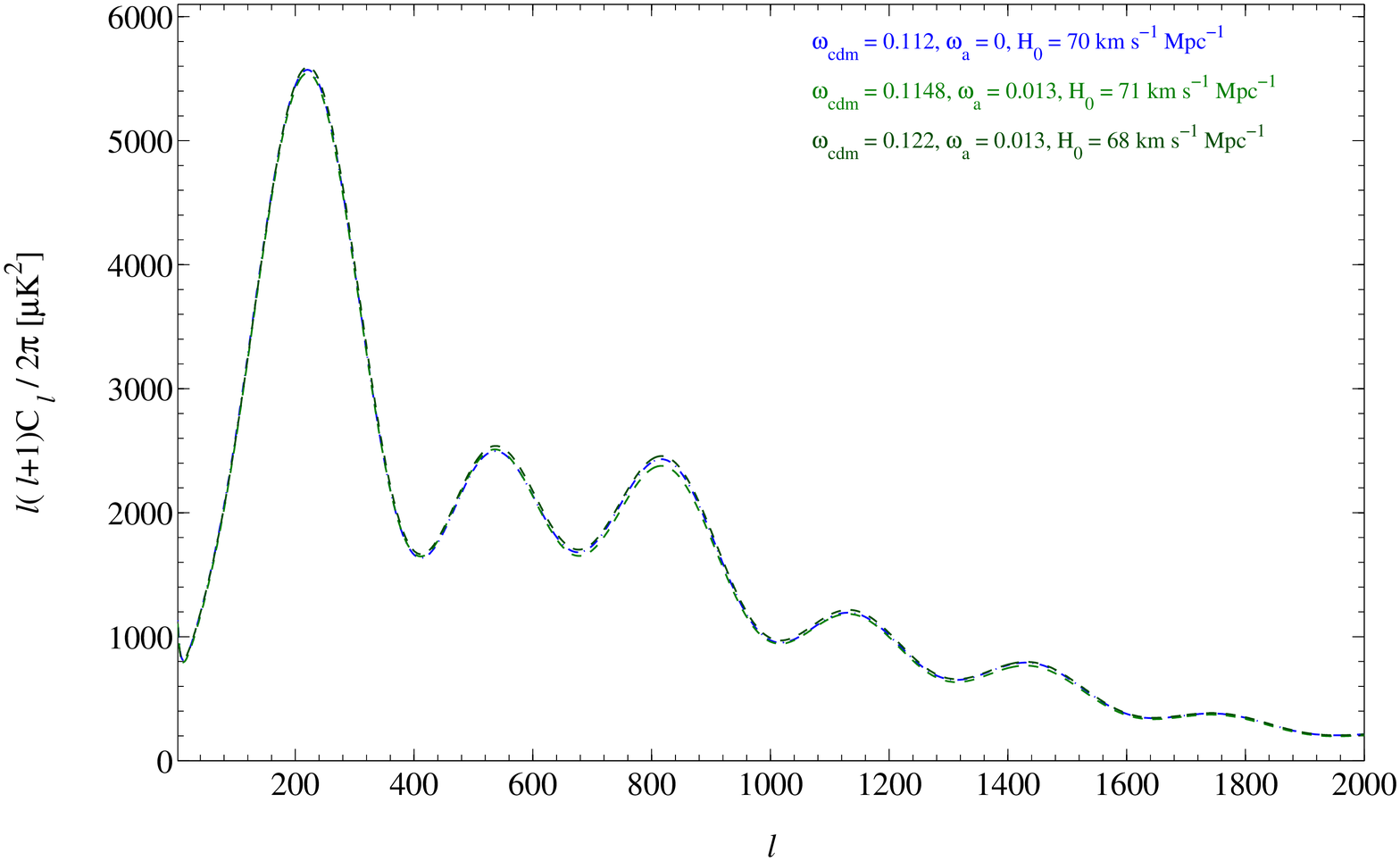}
\caption{CMB temperature angular power spectra for various  cosmological models.  The blue/dot-dash line corresponds to the reference $\Lambda$CDM model ($\omega_{\rm cdm}=0.112, H_0 = 70\  {\rm km} \ {\rm s}^{-1} \ {\rm Mpc}^{-1}$).  In the top panel, the red/solid line denotes the case of three degenerate neutrinos with mass
sum $\Sigma \, m_\nu=1.2$~eV  ($\omega_\nu=0.013$),  the total dark matter density $\omega_{\rm dm}$ raised to $0.1266$, and $H_0=60\  {\rm km} \ {\rm s}^{-1} \ {\rm Mpc}^{-1}$.
In the bottom panel, the green/dashed lines represent two axion scenarios with $m_a=2.4$~eV ($\omega_a=0.013$).  In one case $\omega_{\rm dm}=0.1148$ and $H_0=71 \  {\rm km} \ {\rm s}^{-1} \ {\rm Mpc}^{-1}$, in the second  $\omega_{\rm dm}=0.122$  and $H_0=68 \  {\rm km} \ {\rm s}^{-1} \ {\rm Mpc}^{-1}$.
The primordial fluctuation amplitude has also been adjusted.
}
\label{fig:adjusted}
\end{figure}

This point is illustrated in figure~\ref{fig:rz}, which shows the fractional change in $f(a)$ for a neutrino HDM scenario and an axion scenario of the same present-day energy density, relative to the $\Lambda$CDM case.  To define $f(a)$ we consider to be nonrelativistic all HDM particles with momenta less than a factor $c_{\rm nr}$ of the particle's rest mass, where for $c_{\rm nr}$ we test a range of values from $0.3$ to $1$.  Clearly, independently of our choice of $c_{\rm nr}$,
varying $\omega_{\rm dm}$ only allows the matching of $f(a)$ at one point in time, which we choose here to be $a=a_{\rm eq} \simeq 0.0003$, the scale factor at matter--radiation equality in the reference $\Lambda$CDM model.  Furthermore, the matching demands that we {\it increase} $\omega_{\rm dm}$: by approximately 0.014 in the $\Sigma \, m_\nu=1.2$~eV neutrino case, and by an amount 0.003--0.0205 in the $m_a=1$~eV axion case.  When considered in conjunction with the three-way $(\Sigma \, m_\nu,\omega_{\rm dm},H_0)$-degeneracy [or the $(m_a,\omega_{\rm dm},H_0)$-degeneracy] from $\theta_{\rm s}$ discussed above and in section~\ref{sec:correlation}, this immediately leads us to the following conclusions:

\begin{itemize}

\item For a fixed value of $\Sigma \, m_\nu$, raising $\omega_{\rm dm}$ necessitates that we lower $H_0$ even more in order to match the angular sound horizon of the reference $\Lambda$CDM model.   This reinforces the negative correlation already existing between $\Sigma \, m_\nu$ and $H_0$,  even if no adjustments were made to $\omega_{\rm dm}$ in order to match $f(a_{\rm eq}$).  Indeed, we see in the top panel of figure~\ref{fig:adjusted} that after incrementing $\omega_{\rm dm}$ by an amount 0.0146 and lowering $H_0$ to $60 \ {\rm km} \ {\rm s}^{-1} \ {\rm Mpc}^{-1}$, the resulting CMB temperature power spectrum for the $\Sigma \, m_\nu=1.2$~eV neutrino scenario matches the reference spectrum of the $\Lambda$CDM model with $H_0$ to $70 \ {\rm km} \ {\rm s}^{-1} \ {\rm Mpc}^{-1}$ almost perfectly.

\item For a fixed $m_a$, depending on the chosen $\omega_{\rm dm}$ value,  the correlation between $m_a$ and~$H_0$ induced by fixing the angular sound horizon scale can be mildly positive for no or a small increase in $\omega_{\rm dm}$, or mildly negative at the upper end of the $\omega_{\rm dm}$ range.  Two examples are shown in the bottom panel of figure~\ref{fig:adjusted} for the $m_a=2.4$~eV axion scenario: one case has $\omega_{\rm dm}$ increased by an amount 0.0028 and $H_0$ raised to $71 \ {\rm km} \ {\rm s}^{-1} \ {\rm Mpc}^{-1}$, the other has a 0.01 increment in $\omega_{\rm dm}$ and $H_0=68 \ {\rm km} \ {\rm s}^{-1} \ {\rm Mpc}^{-1}$.  Both cases mimic the reference $\Lambda$CDM spectrum reasonably well in the first two peaks.  Therefore, marginalising over the uncertainty in $\omega_{\rm dm}$ leaves no apparent correlation between $m_a$~and~$H_0$.

Observe also that both adjusted axion scenarios in figure~\ref{fig:adjusted} begin to show similar signs of deviation, although in opposite directions, from the third acoustic peak onwards.  These deviations were not previously measurable by WMAP alone because WMAP was cosmic variance-limited only up to $\ell  \sim 500$. A $m_a=2.4$~eV axion scenario was therefore allowed by WMAP alone.  Planck is able to rule out this scenario simply by virtue of its cosmic variance-limited measurement of the third acoustic peak.

\end{itemize}

Lastly, we briefly discuss the role of the angular diffusion scale $\theta_{\rm d}$ measurement on the neutrino and the axion mass bounds.  To this end, it is instructive to first consider the simpler case of a nonstandard $N_{\rm eff}$; here, a larger $N_{\rm eff}$ requires that we increase both $\omega_{\rm dm}$ and $H_0$ in order to fit the epoch of matter--radiation equality and the angular sound horizon.  However, increased values of $\omega_{\rm dm}$ and $H_0$ also lead to a larger $\theta_{\rm d}$, defined as $\theta_{\rm d} \equiv r_{\rm d}(z_\star)/D_{\rm A}(z_\star)$, where
\begin{equation}
r^2_{\rm d} (z_\star) \simeq (2 \pi)^2 \int_{z_\star}^{\infty} \frac{dz}{a \sigma_T n_e(z) H(z)} \left[ \frac{R^2 + (16/15) (1+R)}{6(1+R)^2} \right]
\end{equation}
is the (squared) photon diffusion scale at decoupling, with $\sigma_T$  the Thomson scattering cross-section, $n_e(z)$ the free electron number density, and
$R (z) \equiv (3/4)(\rho_b/\rho_\gamma)$ the baryon-to-photon energy density ratio.
Thus, the net effect is that a larger $N_{\rm eff}$ causes more damping of the CMB temperature power spectrum at $\ell > 1000$, and measurement of this damping tail can be a very effective means to pin down the value of $N_{\rm eff}$ (see, e.g., \cite{Hou:2011ec,Abazajian:2012ys}).

The same is not true however for the $\Sigma \, m_\nu$ inference.  This is because while we do need to increase $\omega_{\rm dm}$ to match the radiation-to-matter energy density ratio, the $H_0$ value must be simultaneously lowered in order to fit the angular sound horizon scale; these two adjustments more or less cancel each other's effect on $\theta_{\rm d}$.  In comparison, the axion scenario has a marginally larger impact on the angular diffusion scale than the neutrino case because in order to fit $\theta_{\rm s}$, $H_0$ can be held more or less fixed.  Physically, the difference can be traced to the fact that apart from the early ISW effect, the main impact of massive neutrinos that become nonrelativistic in bulk {\it after} photon decoupling on the CMB anisotropies is through the angular diameter distance to the last scattering surface $D_A(z_\star)$.  But since $D_A(z_\star)$ simply rescales the positions of the acoustic peaks and hence alters $\theta_{\rm s}$ and $\theta_{\rm d}$ in the same way, the ratio $\theta_{\rm d}/\theta_{\rm s}$ is independent of $\Sigma m_\nu$.  In contrast, the majority of axions in the mass range considered become nonrelativistic {\it before} photon decoupling.  Such axions therefore have a more direct impact on the physical sound horizon $r_{\rm s}$ and the diffusion scale $r_{\rm d}$ at decoupling, and remain discernible in the ratio $\theta_{\rm d}/\theta_{\rm s}$.  Nonetheless, 
the overall effect of axion masses on $\theta_{\rm d}/\theta_{\rm s}$ is still minor in comparison with $m_a$'s irreducible effects on the radiation-to-matter energy density ratio and consequently the peak height ratios, and is therefore not a decisive factor in cosmological  constraints on $m_a$.


\end{document}